\documentclass[prd,onecolumn,preprintnumbers,amsmath,amssymb,superscriptaddress]{revtex4-1}
\usepackage{bm,curves}
\usepackage{epsfig}
\usepackage{natbib}
\usepackage{dcolumn}
\usepackage{lineno}
\usepackage{amsmath}
\usepackage{slashed}
\usepackage{xcolor}
\usepackage{subfigure}
\usepackage[utf8]{inputenc}
\modulolinenumbers[5]

\def\lan{\langle }
\def\ran{\rangle }
\def\ksl{\slashed{k} }
\newcommand{\sigc}{\sigma_{c\overline{c}}}
\newcommand{\fcc}{F^{c\overline{c}}_{2,\, \mathrm{IC}}}

\begin{document}

\preprint{NT@UW-17-09}

\title{A Bayesian analysis of light-front models and the nucleon's charmed sigma term}

\author{T. J. Hobbs}
\email{tjhobbs@uw.edu}
\affiliation{
        Department of Physics,
         University of Washington, Seattle, Washington 98195, USA
}
\author{Mary Alberg}
\email{alberg@phys.washington.edu}
\affiliation{
        Department of Physics,
         University of Washington, Seattle, Washington 98195, USA
}
\affiliation{
        Department of Physics,
         Seattle University,
         Seattle, Washington 98122, USA
}
\author{Gerald A.~Miller}
\email{miller@uw.edu}
\affiliation{
        Department of Physics,
         University of Washington, Seattle, Washington 98195, USA
}

\begin{abstract}
We present the results of a recent analysis to study the nucleon's charm sigma term, $\sigc$. We construct
a minimal model in terms of light-front variables and constrain the range of possibilities using extant
knowledge from deeply inelastic scattering (DIS) and Bayesian parameter estimation, ultimately computing $\sigc$
in an explicitly covariant manner. We find a close correlation between a possible nonperturbative component of the
charm structure function, $\fcc$, and $\sigc$. Independent of prescription for the covariant
relativistic quark-nucleon vertex, we determine $\sigc$ under several different scenarios for the magnitude of
intrinsic charm (IC) in DIS, namely $\lan x \ran_{c+\overline{c}} = 0.1\%$, $0.35\%$, and $1\%$, obtaining for
these $\sigc = 4 \pm 4$, $12 \pm 13$, and $32 \pm 34$  MeV, respectively. These results imply the existence of a
reciprocity between the IC parton distribution function (PDF) and $\sigc$ such that new information from either DIS or improved
determinations of $\sigc$ could significantly impact constraints to the charm sector of the proton wave function.
\end{abstract}


\date{\today}
\maketitle
%
\section{Introduction}
\label{sec:intro}
\begin{figure}
\label{fig:1a}
\includegraphics[scale=0.5]{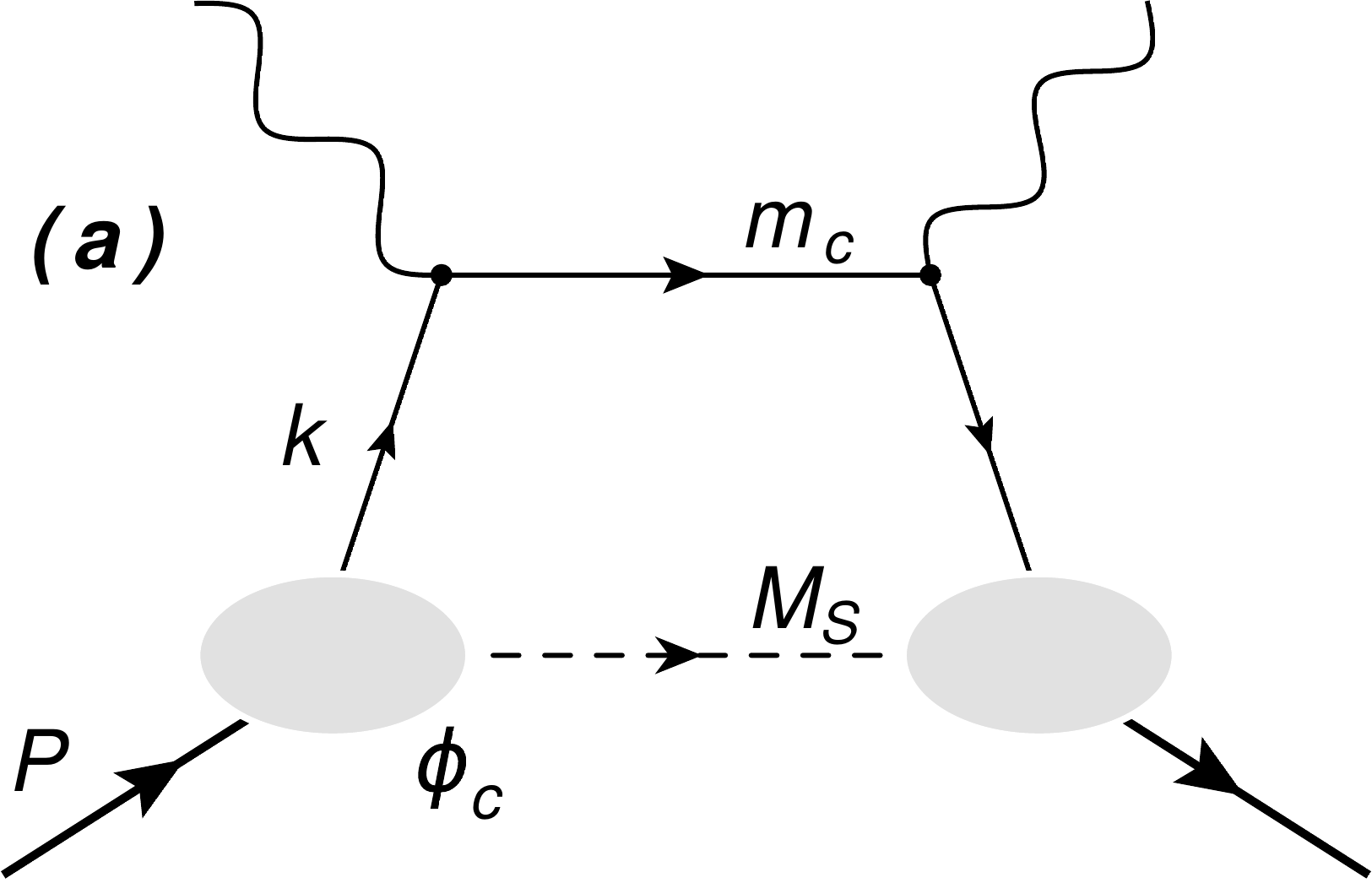} \ \ \ \ \ \ \ \ \ \ \ \
\includegraphics[scale=0.5]{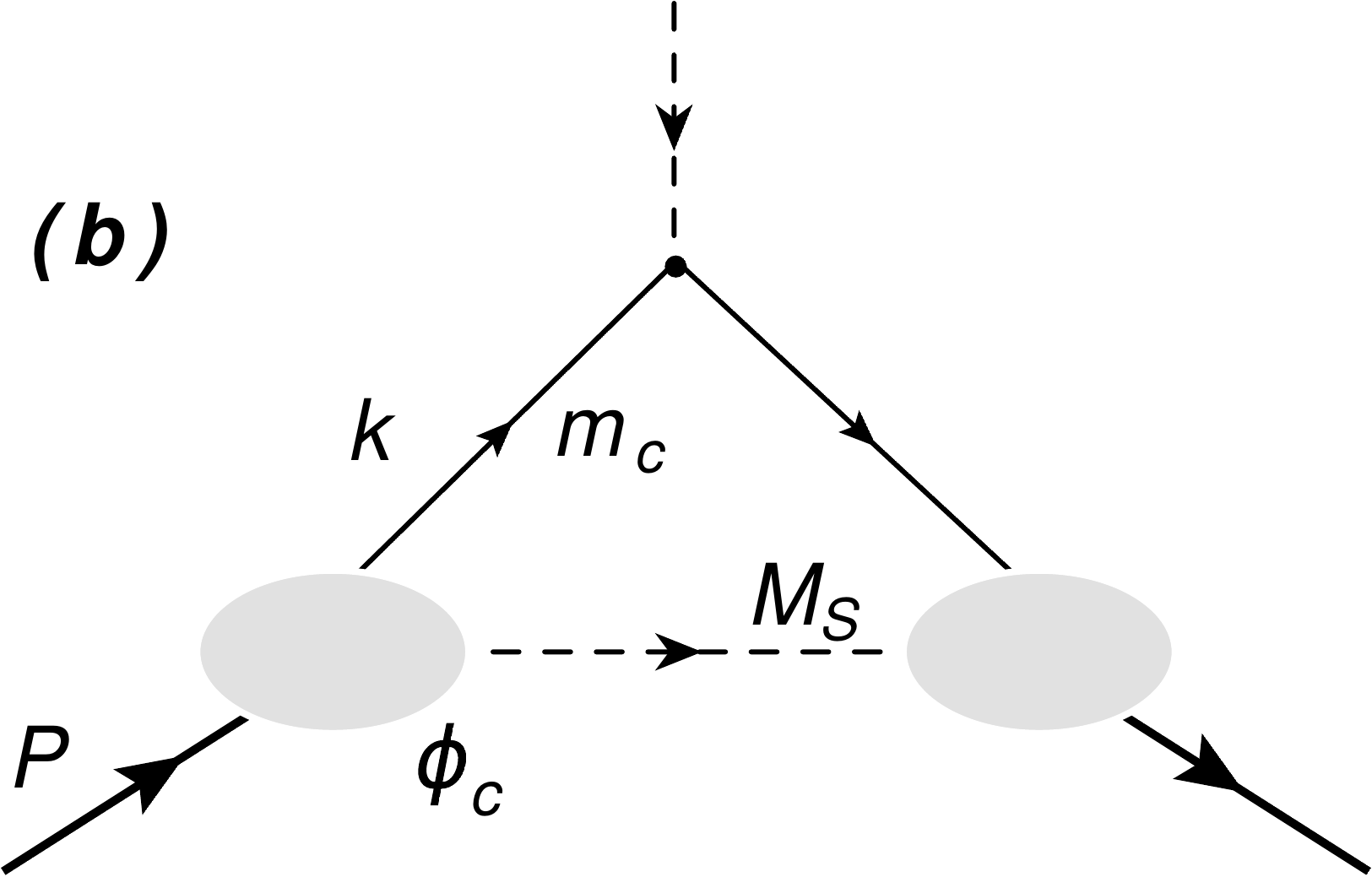}
\label{fig:1b}
\caption{(Color online) ({\bf a}) The leading contribution to the nonperturbative charm structure function from scattering off the $5$-quark Fock
state represented by Eq.~(\ref{eq:Fock}) in the scalar spectator model. An external photon interacts with a charm quark of momentum
$k$, which emerges from a dissociation of the proton (momentum $P$) with a recoiling $4$-quark scalar
spectator state of static mass $M_S$. A complementary graph in which the anti-charm quark is struck
(with an analogous $4$-quark $qqqc$ spectator with a mass we denote $M_{\bar{S}}$) is not shown. ({\bf b}) The
corresponding graph responsible for the nucleon's charm sigma term interpreted as a scalar form factor as in
Eqs.~(\ref{eq:sigFORM})--(\ref{eq:sigCOV}). In both panels, the oval blobs labeled $\phi_c$ at the quark-nucleon
vertices correspond to the phenomenological vertex factor defined in Eq.~(\ref{eq:phi}).
}
\label{fig:diagram}
\end{figure}

\paragraph{Motivation.}
Understanding how fundamental QCD constituents generate the nucleon's bulk properties --- its mass, spin, and effective interactions
with external electroweak probes --- has been a constant theme in hadronic physics since the field's inception. Of these, the origin
of the proton mass $M$ is arguably the most basic, and great strides have recently been made computing $M$ using nonperturbative
techniques \cite{Leinweber:2003dg,Procura:2003ig,AliKhan:2003ack}. At the same time, a complete reckoning of the various QCD effects
that go into $M$ remains an unrealized goal, and the possible role of virtual heavy quark states persists as an unresolved issue.
In this context sigma terms \cite{Cheng:1970mx,Thomas:2001kw} have a long history as a convenient lever quantifying the contribution
of specific states to the nucleon mass. Going beyond $SU(2)$, the charm quark contribution to the nucleon's sigma term is defined as
the matrix element of a scalar quark-level density operator
\begin{equation}
\sigc\ \equiv\ m_c\, \langle p \big|\, \overline{c} c\, \big| p \rangle\ .
\label{eq:sigdef}
\end{equation}
As such, it represents a generalization to the heavy-quark sector of the pion-nucleon and strange sigma terms
\cite{Junnarkar:2013ac,Alexandrou:2013nda,Ren:2014vea,Yang:2015uis,Durr:2015dna} interpreted as describing the origin of the nucleon's
mass from the current masses of its partonic constituents. This quantity is not directly accessible to experiment, but is calculable
using a number of nonperturbative methods, including lattice QCD, which has produced the values
\cite{Freeman:2012ry,Gong:2013vja,Abdel-Rehim:2016won}
\begin{align}
\sigc\ &=\ 73\ (35)\,\mathrm{MeV}\ \ \ \ \ \ \ \ \ \ \ \ \ \ \ \, \mathrm{Ref.}\ [11] \nonumber\\
&=\ 94\ (31)\,\mathrm{MeV}\ \ \ \ \ \ \ \ \ \ \ \ \ \ \ \, \mathrm{Ref.}\ [12] \nonumber\\
&=\ 79\ (21)\binom{12}{8}\,\mathrm{MeV}\ \ \ \ \ \ \ \ \mathrm{Ref.}\ [13]\ ,
\end{align}
where in the last line for Ref.~\cite{Abdel-Rehim:2016won} the separated uncertainties are statistical and systematic, respectively.
Consistent higher-order pQCD calculations have also been explored out to $\mathcal{O}(\alpha^3_s)$ \cite{Kryjevski:2003mh,Shifman:1978zn},
while, more recently, a calculation based on the chiral constituent quark model \cite{Duan:2016rkr} obtained a somewhat lower magnitude.
Beyond its significance to the theory of proton structure, the nucleon's charm sigma term may play a potentially important
role in searches for new physics beyond the Standard Model (BSM) --- especially direct searches for weakly interacting
massive particles (WIMPs) \cite{Jungman:1995df,Cline:2013gha}. Generically, direct dark matter searches are sensitive to
putative WIMP-nucleon interactions in the form of, {\it e.g.}, hypothetical neutralino-nucleus scattering \cite{Agnese:2014aze,Akerib:2013tjd}.
Standard Model degrees of freedom might see this dark sector by exchanging a spin-$0$ Higgs boson, which interacts
with the nucleon as depicted in Fig.~\ref{fig:diagram}({\bf b}); the strength of the WIMP-nucleon effective coupling
is then closely related to the quark fields' scalar densities as given by their sigma terms. This connection has been explored
systematically by operator analyses of WIMP-nucleon scattering as in the recent study of Hill and Solon \cite{Hill:2013hoa}.
Approaches of this type generally rely on heavy-quark effective theory to integrate away the heaviest electroweak scales such that the
leading spin-$0$ contribution to the interaction of light, neutral WIMPs ($\chi_\nu$) with the nucleon's partonic degrees of freedom is
schematically
\begin{equation}
\mathcal{L}_{\chi_\nu}\ =\ {1 \over M^3_W} \overline{\chi}_\nu \chi_\nu \left( \sum_q\, c^{(0)}_q \big( m_q \overline{q}q \big)
+ c^{(0)}_g \big( G^A_{\mu\nu} \big)^2 \right)\ +\ \dots\ ,
\end{equation}
in which the dots above include spin-$2$ matrix elements and higher-order contributions suppressed by larger powers of $1 \big/ M_W$.
In this formalism the Wilson coefficients $c^{(0)}_{q,g}$ can be computed to a given order, and a WIMP-nucleon cross section
determined as a function of the nucleonic matrix element of $\sigma_q \equiv m_q \overline{q}q$; as demonstrated in
FIG.~$3$ of Ref.~\cite{Hill:2013hoa}, the contribution of heavy quarks --- especially charm --- can be particularly important, with
$\sigc$ playing a decisive role in determining the strength of the WIMP-nucleon interaction in some BSM scenarios involving extended
electroweak sectors.
In general, recent determinations \cite{Freeman:2012ry,Gong:2013vja,Abdel-Rehim:2016won,Kryjevski:2003mh,Duan:2016rkr} give
values $25 \lesssim \sigc \lesssim 125$ MeV \cite{Hill:2013hoa}, for which heavy-particle effective theory suggests an upper limit
for the spin-independent WIMP-nucleon cross section $\sigma_{\rm SI} \lesssim 10^{-48}$ cm$^2$;
for $\sigc \lesssim 25$ MeV, however, a spin-independent cross section $\sigma_{\rm SI} \sim 10^{-49}$ cm$^2$
is {\it predicted} in the electroweak doublet scenario \cite{Hill:2013hoa}. For this reason, improved
knowledge of $\sigc$ would provide useful guidance to ongoing direct WIMP searches. 
Parallel to these recent developments, the proton's intrinsic charm (IC) content --- a concept typically rooted
in QCD global analyses of nucleon parton distribution functions (PDFs) --- remains a controversial subject, with unambiguous
phenomenological evidence for its existence remaining elusive. When coupled with the fact that the earlier-mentioned information
from lattice QCD on $\sigc$ is relatively sparse and dominated by large uncertainties, the present status of IC PDF studies suggests
that it would be invaluable to formulate an overarching framework to comprehensively analyze nucleon charm in both sectors; once
constrained to experimental inputs, the resulting formalism might shed light on charm in the nucleon wave function by allowing
information on the IC PDF to reciprocally improve knowledge of $\sigc$, and {\it vice versa}. Such an analysis might be enhanced
further by contemporary advances in computational abilities, which have made resource-intensive methods founded in Bayesian
inference like Markov-chain Monte Carlo (MCMC) a more common and powerful approach. While Bayesian techniques have enabled progress
in a number of fields --- {\it e.g.}, chiral EFTs for nuclear structure \cite{Zhang:2015ajn} and extractions of QCD equations of
state and bulk properties in heavy-ion collisions \cite{Sangaline:2015isa} --- they have seldom been deployed in model analyses
of hadronic structure; for this reason, we demonstrate a representative calculation in the present article. In particular, we show
how MCMC numerical methods may constrain a {\it minimal} model for the charm contribution to the nucleon's structure
function and sigma term, assuming for specificity hypothetical scenarios for the IC PDF/structure function and systematically
evaluating the resulting predictions for $\sigc$. The ultimate goal is the establishment of a connection between the shapes
and moments of IC PDFs as predicted in various recent QCD analyses
\cite{Dulat:2013hea,Jimenez-Delgado:2014zga,Jimenez-Delgado:2015tma,Ball:2016neh,Rojo:2017xpe,Hou:2017khm} and the magnitude of
the charm sigma term, as we expect this connection may provide leverage on both heretofore independent lines of inquiry.
Noting that detailed empirical information on the nonperturbative charm structure function is relatively lacking, for our underlying
framework we adopt in Sect.~\ref{sec:LF} a simple picture in which the nucleon's charm content resides in a leading $5$-quark
$\big|uudQ\overline{Q}\ran$ Fock state, wherein the interacting charm quark is dressed by a scalar spectator of fixed mass.
More elaborate physical choices can be made for the interacting structure of the nonperturbative $5$-quark state, but inevitably
lead to the introduction of additional fitting parameters; while these additional parameters will have to be confronted in the
future as higher precision data become available, they are not presently useful given the lack of data to constrain them.
Rather, we will instead constrain the simplified quark $+$ scalar model
with the understanding that dynamics present in more detailed models are effectively absorbed into the tunable parameters
of the present framework. With our quark-spectator picture in hand, we use light-front wave functions (LFWFs) in
Sect.~\ref{ssec:DIS} to compute the IC PDF resulting from the deeply inelastic scattering (DIS) of an external
photon off the $\overline{c}c$ $5$-quark state.
Using the LFWFs as a vehicle to constrain the parameter space of the quark-spectator model from hypothetical pseudo-data
inspired by effective models of the IC PDF \cite{Hobbs:2013bia}, we then compute the scalar form factor pictured in
Fig.~\ref{fig:diagram}({\bf b}) with standard covariant techniques, leading to an explicit evaluation of $\sigc$
for a wide range of model parameters as desired. We lay out this calculation in Sect.~\ref{ssec:sigmodel}, and describe the
outline of the MCMC procedure in Sect.~\ref{sec:MCMC}, summarizing the numerical results for $\sigc$. We conclude
in Sect.~\ref{sec:conc} by assessing the established connection between the IC PDF and $\sigc$, characterizing
the resulting implications for the ongoing search for intrinsic charm, and recommending possible phenomenological avenues.
%

%
%
%
%
\section{quark $+$ scalar model}
\label{sec:LF}
We proceed with the picture that the leading charm contribution to the nucleon's nonperturbative
wave function is realized through a $\big|uudc\overline{c}\ran$ state in which a virtual charm quark
is dressed by a $4$-quark scalar. This configuration contributes to the proton's charm structure
function and sigma term, respectively, through the diagrams sketched in Figs.~\ref{fig:diagram}({\bf a}) and
\ref{fig:diagram}({\bf b}). This physical picture has antecedents in treatments of the nucleon's
strange \cite{Hobbs:2014lea,Hobbs:2016xlg} and bare valence quark content \cite{Cloet:2012cy}, and in both cases
LFWFs were an advantageous means of computing the quark-level PDFs. We extend this framework to model
IC in DIS in Sect.~\ref{ssec:DIS} below, but must supplement our formalism with an explicitly covariant
scheme to adapt the model to $\sigc$.
That this is necessary is due to long-known properties of the light-front formalism~\cite{Chang:1973qi,Lepage:1980fj}.
For $k^+ > 0$, quantizing a field theory of spin-$1/2$ particles on the light front leads to a set of Feynman rules
that specify the fermion light-front propagator as
\begin{equation}
i{\slashed{\overline{k}} + m \over k^2 - m^2 + i\epsilon}\
=\ i{\sum_\lambda u_\lambda(k)\, \overline{u}_\lambda(k) \over k^2 - m^2 + i\epsilon}\
=\ {i \over \ksl - m + i\epsilon}\, -\, i{\gamma^+ \over 2k^+}\ ,
\label{eq:LFprop}
\end{equation}
such that the resulting $3$-dimensional LFWFs, which depend upon the on-shell ($\overline{k}{}^2=m^2$)
quark momentum $\overline{k}{}^\mu$, involve contributions from noncovariant, or ``instantaneous'',
pieces of the light-front propagator $\sim\! \gamma^+$. The covariance of the LFWF formalism is restored
by the fact that matrix elements relevant for the electromagnetic processes of usual interest --- typically,
DIS and elastic scattering --- consist of operator products with overall factors of $\gamma^+$ which
void the offending noncovariance by the identity $\gamma^+\! \gamma^+ = 0$.
While this approach ensures the covariance of PDF calculations as in Eq.~(\ref{eq:cx}) of Sect.~\ref{ssec:DIS}
below, the charm sigma term $\sigc$ is the matrix element of a scalar operator; as such, directly evaluating
Eq.~(\ref{eq:sigdef}) in terms of LFWFs would be approximative due to the presence of the unsubtracted noncovariant
contributions in Eq.~(\ref{eq:LFprop}). We rectify this by instead computing $\sigc$ in Sect.~\ref{ssec:sigmodel} as an
explicitly covariant amplitude according to the diagram in Fig.~\ref{fig:diagram}({\bf b}). We may then evaluate $\sigc$
using the quark $+$ scalar model parameters constrained by MCMC simulations of $x(c+\overline{c})$, which {\it is}
covariantly calculable from LFWFs. We summarize the numerical results of this procedure in Sect.~\ref{sec:MCMC}.
%
%
\subsection{DIS sector}
\label{ssec:DIS}
\paragraph{Collinear PDF.}
LFWFs provide a systematic, covariant description of the nucleon's partonic substructure as imaged in external electromagnetic
interactions and have recently been used to analyze the strange component of the proton's form factors
\cite{Hobbs:2014lea,Hobbs:2016xlg}. Following this earlier work, the leading contribution to the proton's charm content in DIS can be
obtained in analogous fashion by expanding its wave function in terms of a $5$-quark state, for which the leading component is taken
to consist of a struck charm quark/antiquark accompanied by a recoiling $4$-quark scalar spectator as shown in Fig.~\ref{fig:diagram}({\bf a}):
\begin{equation}
|\Psi^\lambda_P(P^+,{\boldsymbol P_\perp}) \ran\ =\ {1 \over 16\pi^3} \sum_{q=c,\overline{c}}\int {dx\,d^2\boldsymbol{k_\perp} \over \sqrt{x (1-x)}}\
\psi^\lambda_{q \lambda_q}(x, {\boldsymbol k_\perp})\, |q; xP^+, x{\boldsymbol P_\perp} + {\boldsymbol k_\perp} \ran\ .
\label{eq:Fock}
\end{equation}
We may use this expression and our knowledge of the quark + scalar spectator helicity wave functions
$\psi^\lambda_{q \lambda_q}(x, {\boldsymbol k_\perp})$ from Ref.~\cite{Hobbs:2014lea} to compute the nonperturbative charm structure function
in the quark-parton model
\begin{align}
\label{eq:SF}
\fcc(x,Q^2=m^2_c) &\equiv {4x \over 9}\, \Big(\, c(x) + \overline{c}(x)\, \Big) \\
c(x)\ &=\ {1 \over 16 \pi^2} \int {dk^2_\perp \over x^2 (1-x)}  \Big[{k^2_\perp + (m_c + x M)^2 \over (M^2 - s_{cS})^2 }\Big]\
\big|\phi_c (x,k^2_\perp) \big|^2\ ,
\label{eq:cx}
\end{align}
where $s_{cS}$ is the charm quark-spectator invariant mass given by
\begin{equation}
s_{cS}\left(x,k^2_\perp\right)\ =\ { 1 \over x\, (1-x)} \Big(\, k^2_\perp +  (1-x)\,m^2_c  + x\,M^2_S\, \Big)\ .
\label{eq:s-inv}
\end{equation}
In the expression in Eq.~(\ref{eq:cx}) for the IC PDF, we choose a parametric form for the transverse momentum
dependent quark-nucleon interaction $\phi_c$; this has the physical meaning of resumming an infinite tower of nonperturbative QCD
interactions in the infrared and implementing the dynamics of confinement phenomenologically; practically, $\phi_c$ regulates
the ultraviolet divergences that naturally enter at large $k_\perp$ from the graphs of Fig.~\ref{fig:diagram}. As noted
at the head of Sect.~\ref{sec:LF}, we will compute $\sigc$ using covariant techniques in Sect.~\ref{ssec:sigmodel}, and must for consistency
take an explicitly Lorentz-invariant form for the relativistic quark-nucleon vertex function. A natural choice with a known
history in analyses of pion-nucleon interactions \cite{Alberg:2012wr,McKenney:2015xis} and quark-diquark models
\cite{Melnitchouk:1994en} is a $t$-dependent power-law expression of the form
\begin{align}
\phi_c (x,k^2_\perp)\ =\
\sqrt{g_c}\, \left( {\Lambda^2_c \over t_c - \Lambda^2_c } \right)^\gamma
\label{eq:phi}
\end{align}
for which we generally take $\gamma = 3$ as appropriate for a $4$-parton spectator state, with $t_c$ being the
covariant virtuality of the intermediate charm quark given explicitly by
\begin{equation}
t_c\left(x,k^2_\perp\right)\ =\ {1 \over 1-x}\, \Big(\! -\! k^2_\perp + x \big[ (1-x) M^2 - M^2_S \big]\, \Big)\ .
\label{eq:t}
\end{equation}
Expressions similar to Eqs.~(\ref{eq:cx})--(\ref{eq:t}) hold for $\overline{c}(x)$, but using corresponding choices for the
anti-quark wave function, $\Lambda_c \to \Lambda_{\overline{c}}$, {\it etc}. We note also that scenarios for the behavior of
the nucleon's charm content are often \cite{Dulat:2013hea,Jimenez-Delgado:2014zga,Jimenez-Delgado:2015tma} delimited in
terms of the total charm momentum or PDF first moment at partonic threshold ($Q^2=m^2_c$), which can be cast in terms of
the IC structure function of Eq.~(\ref{eq:SF}) as
\begin{equation}
\lan x \ran_{c+\overline{c}}\ \equiv\ {1 \over e^2_c} \int_0^1 dx\, \fcc(x,Q^2=m^2_c)
\end{equation}
(which should not be confused with the distribution mean of $x$); in Sect.~\ref{sec:MCMC} we will use this to constrain MCMC
simulations for $\sigc$.
%

\begin{figure}
\hspace{-0.4cm}
\includegraphics[scale=0.32]{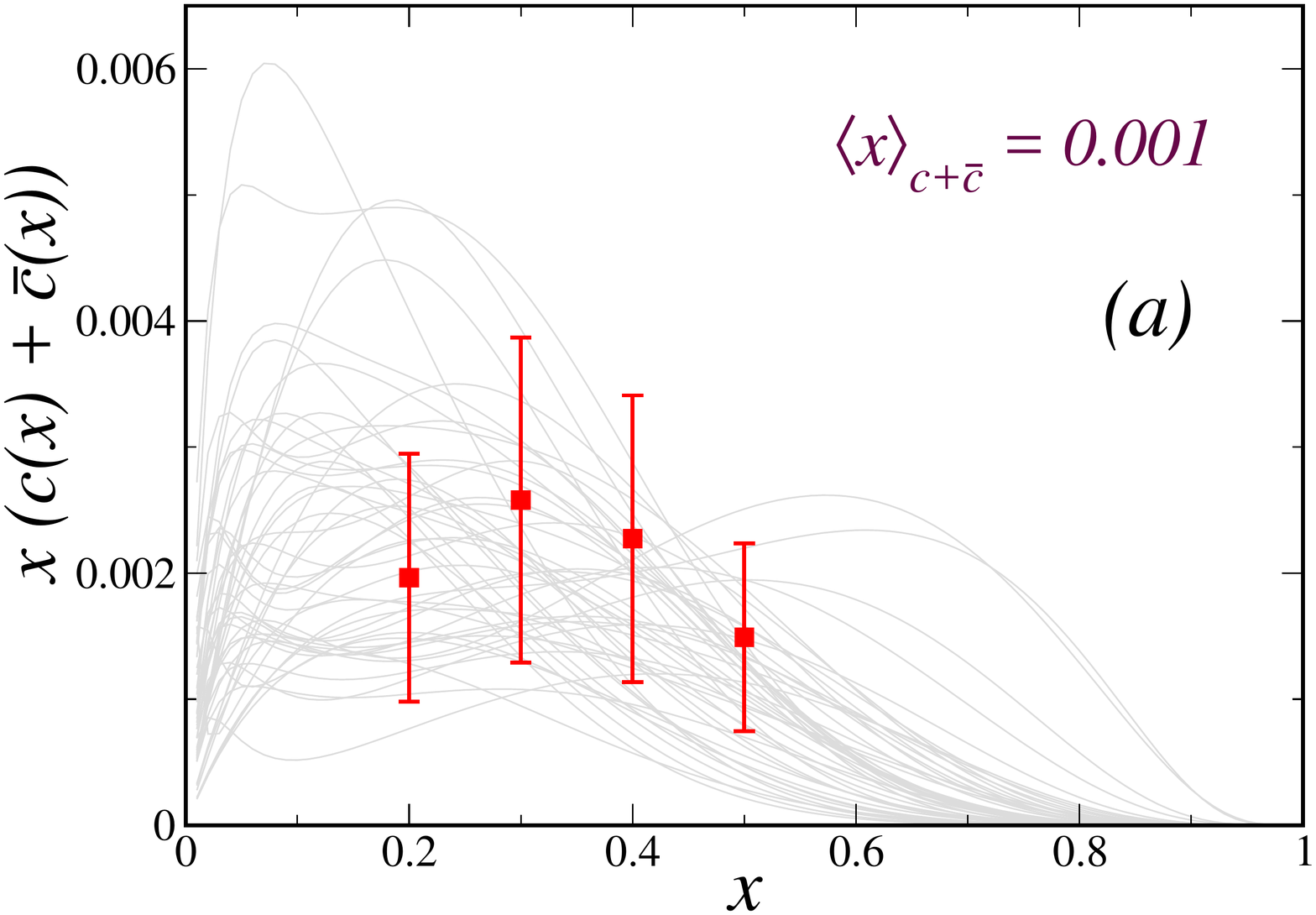} \ \
\includegraphics[scale=0.32]{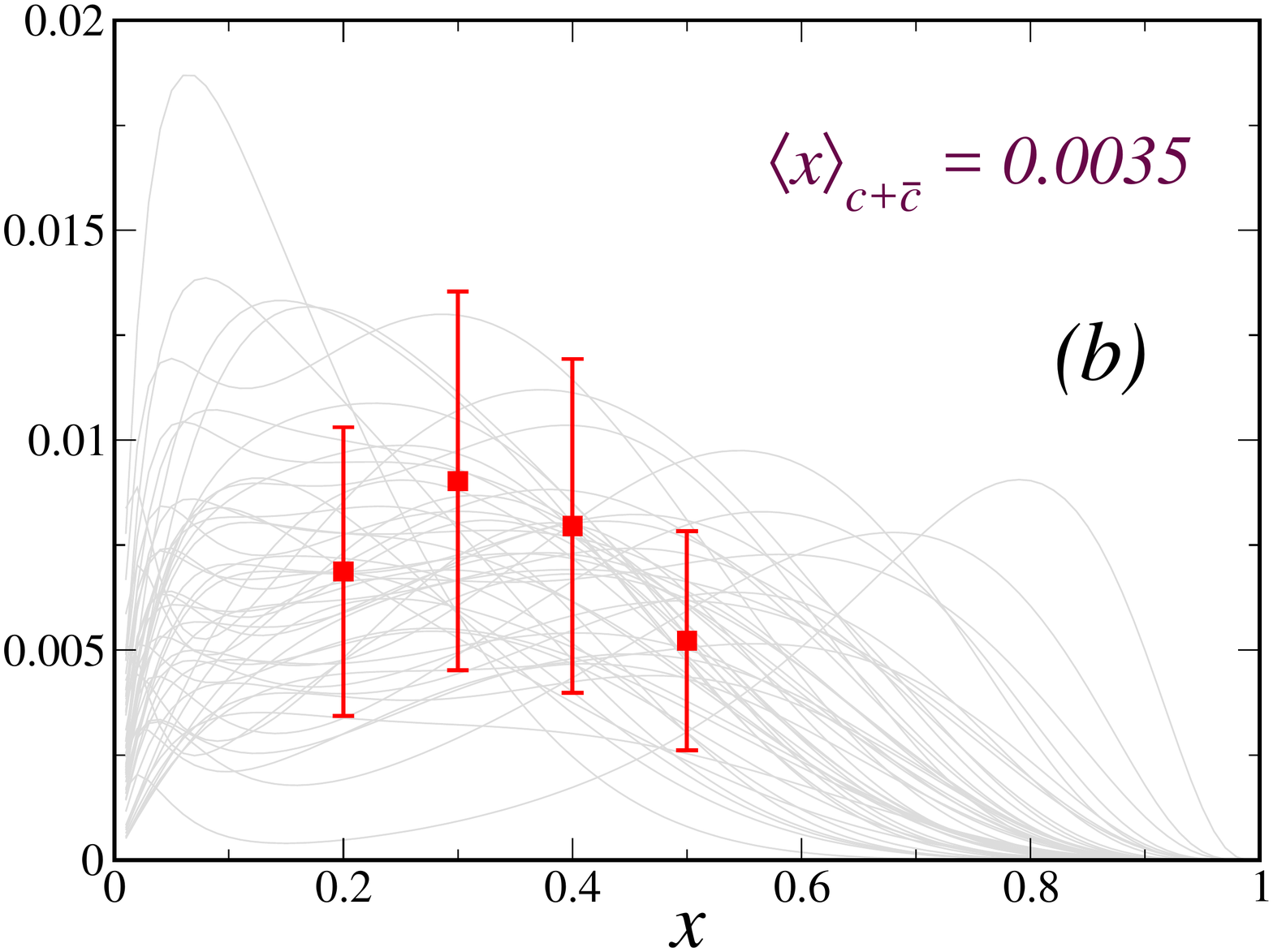}
\caption{
(Color online) Bayesian MCMC runs are constrained with pseudo-data generated by a typical model \cite{Hobbs:2013bia} for
$x[c(x)+\overline{c}(x)]$ to correspond to $\lan x \ran_{c+\overline{c}} = 0.001$ ({\bf a}), $0.0035$ ({\bf b}), and $0.01$ (not
shown here) up to a conservative $50\%$ uncertainty. Alongside this information, we plot model predictions for $x[c(x)+\overline{c}(x)]$
using the covariant power-law vertex function with $\gamma=3$ (gray curves) produced by a small, random sampling of the ensemble of
points in parameter space represented by the chain generated by completed MCMC simulations.
}
\label{fig:LF-model}
\end{figure}
\begin{figure}
\hspace*{-0.25cm}
\includegraphics[scale=0.8]{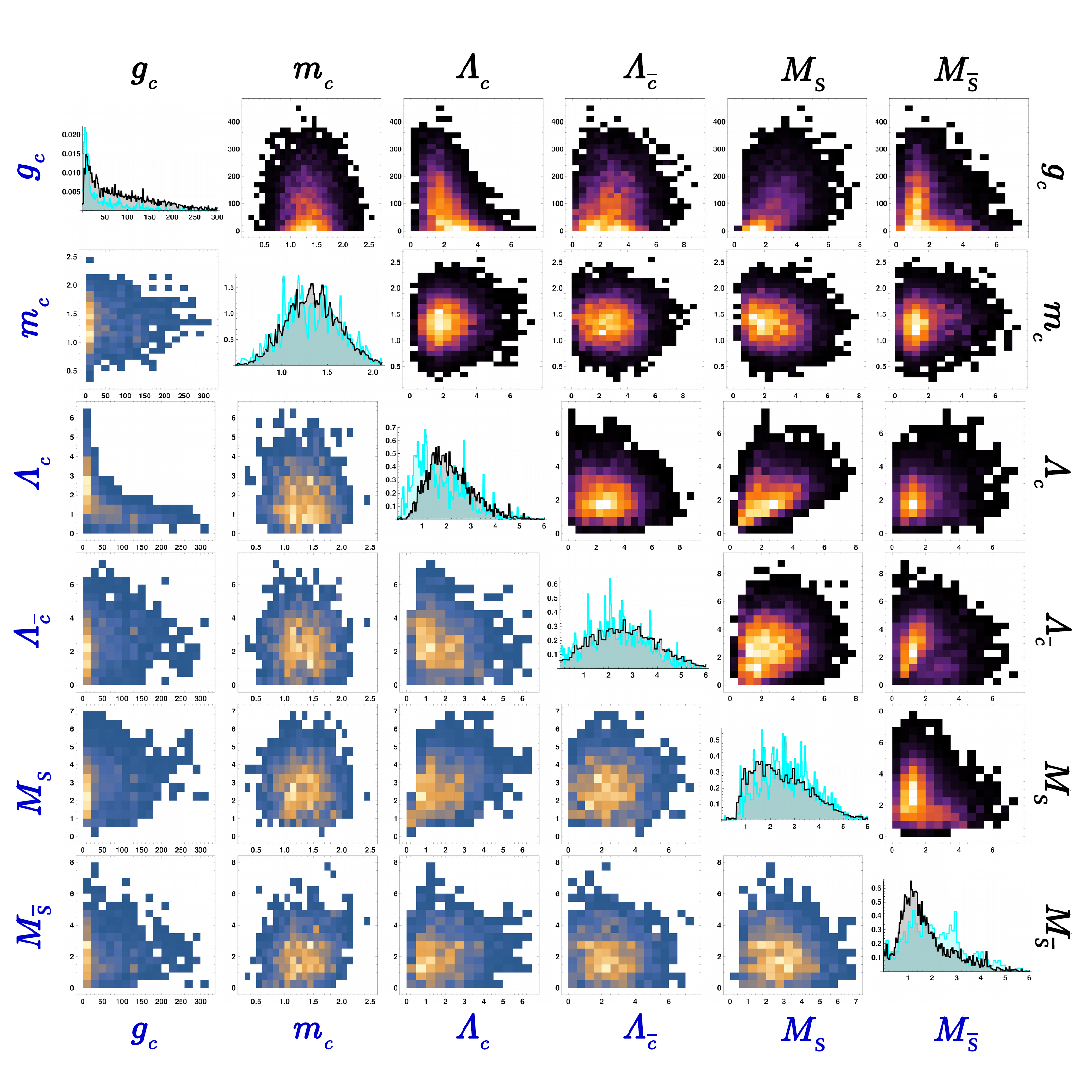}
\vspace*{-1cm}
\caption{
(Color online) MCMC chain statistics for the projected joint posterior distribution of the $6$-parameter model constrained by the DIS pseudo-data
shown in Fig.~\ref{fig:LF-model}({\bf b}), which have the intermediate normalization $\lan x \ran_{c+\overline{c}} = 0.0035 \pm 50\%$. From left-to-right and
top-to-bottom, panels correlate $g_c$,
$m_c$, $\Lambda_c$, $\Lambda_{\bar{c}}$, $M_S$, and $M_{\bar{S}}$. Diagonal entries plot normalized histograms obtained with either the
$\gamma = 1$ (blue) or $\gamma = 3$ (black) charm quark-nucleon interaction of Eq.~(\ref{eq:phi}) for each of these parameters (effectively, the
parameters' marginalized posterior distributions), showing the number of occurrences of specific values over $n_{sim} = 10^5$ runs. Off-diagonal
elements yield the correlations among different parameters of the model, which tend to be relatively weak, especially for the spectator masses
$M_{S,\bar{S}}$; correlations plotted below the diagonal entries are found using the weaker vertex function with the $\gamma = 1$ power law
in Eq.~(\ref{eq:phi}), while the top-right off-diagonal panels correspond $\gamma = 3$ in Eq.~(\ref{eq:phi}). In several cases, there are more
pronounced correlations --- for instance, between $g_c$ and $\Lambda_c$ as well as between $\Lambda_c$ and $M_S$.
}
\label{fig:2D}
\end{figure}

Our choice of $\gamma=3$ in Eq.~(\ref{eq:phi}) is motivated by the physical requirement that the threshold behavior of the resulting
PDFs conform to counting rules \cite{Drell:1970xz,West:1970av}, which dictate $q(x) \sim (1-x)^{2 n_s - 1}$ for $ x \to 1$, where
$n_s$ is the effective number of partonic spectators to the quark-photon interaction. The choice $\gamma=3$ then ensures that the
high-$x$ behavior of the light-front PDFs is $\sim\! (1-x)^7$, consistent with a $4$-quark spectator state. In fact, this behavior
may be verified directly by evaluating the expression
\begin{equation}
n_s\ =\ {1\over2}\, \Big( 1\ -\ \lim_{x\to1}\, (1-x) {d \over dx} \log q(x) \Big)
\label{eq:counting}
\end{equation}
on a large ensemble of quark distributions produced with the numerical techniques described in Sect.~\ref{sec:MCMC} below. As a typical
example, using $\gamma = 3$ we obtain $n_s = 3.99 \pm 0.08$ for the central normalization of this study. Other choices are also possible
(in general, $n_s \approx \gamma + 1$ for $\gamma \ge 1$ such that, {\it e.g.}, a monopole with $\gamma=1$ results in $n_s \approx 2$
consistent with a bare valence quark), but ultimately give qualitatively very similar results. We illustrate the resulting behavior
in constrained model parameters with $\gamma=1$ alongside $\gamma=3$ in Fig.~\ref{fig:2D}.
%
%
%
%
\paragraph{TMDs.}
By construction the collinear PDF of Eq.~(\ref{eq:cx}) may be interpreted as originating from the moment of an unintegrated, transverse momentum
dependent (TMD) function we designate $f_{c/p}(x, \boldsymbol{k}_\perp)$, which is related to $c(x)$ of Eq.~(\ref{eq:cx}) and an analogous
${\boldsymbol k}_\perp$ distribution $f_c({\boldsymbol k}_\perp)$ via \cite{Collins:2011zzd}
\begin{align}
\label{eq:TMDpdf}
c(x)\ &=\ \int d^2 \boldsymbol{k}_\perp\, f_{c/p}(x, \boldsymbol{k}_\perp) \\
f_c({\boldsymbol k}_\perp)\ &=\ \int dx\, f_{c/p}(x, \boldsymbol{k}_\perp)\ ,
\label{eq:TMD}
\end{align}
such that the total charm quark number (zeroth moment of the IC PDF) is given by
\begin{equation}
n_c\ =\ \int dx\, d^2 \boldsymbol{k}_\perp\, f_{c/p}(x, \boldsymbol{k}_\perp)\ = \int dx\, c(x)\ ;
\end{equation}
momentum sum rules require $n_c = n_{\bar{c}}$, a condition we use to fix $g_{\bar{c}}$. Since the matrix elements in this study are in the forward
limit [$\lan p'| \cdots | p \ran$ with $p'=p$], angular integrals are trivial as in Eq.~(\ref{eq:cx}) and the ensemble-averaged
$k_\perp \equiv |{\boldsymbol k}_\perp|$ is then defined by
\begin{equation}
\lan k_\perp \ran_c\ \equiv\ {1 \over n_c} \int dx\, dk^2_\perp\, k_\perp\, f_{c/p}(x, k_\perp)\ ,
\label{eq:kt1}
\end{equation}
and we take the total average charm transverse momentum $\lan k_\perp \ran_{c+\overline{c}}$ to be the mean of separate quark and antiquark
contributions:
\begin{equation}
\lan k_\perp \ran_{c+\overline{c}}\ \equiv\ {1 \over 2}\, \Big( \lan k_\perp \ran_c\, +\, \lan k_\perp \ran_{\overline{c}} \Big)\ .
\label{eq:kt2}
\end{equation}

The TMD-averaged $\lan k_\perp \ran_{c+\overline{c}}$ is linked to the quark distribution in transverse impact parameter space (with
${\boldsymbol k}_\perp$ conjugate to the variable ${\boldsymbol b}_\perp$), a fact which establishes the connection between the transverse momentum
dependence in the light-front model and the physical distributions of charm quarks within the nucleon as part of its transverse tomography.
Formally, a ${\boldsymbol b}_\perp$-dependent distribution can be defined by taking the appropriate Fourier transform of the unintegrated
quark TMD density $f_{c/p}(x,{\boldsymbol k}_\perp)$ of Eqs.~(\ref{eq:TMDpdf})--(\ref{eq:TMD}),
\begin{equation}
f_{c/p}(x, \boldsymbol{b}_\perp)\ =\ \int {d^2 \boldsymbol{k}_\perp \over (2\pi)^2}\, e^{-i {\boldsymbol k}_\perp\! \cdot {\boldsymbol b}_\perp}\,
f_{c/p}(x, \boldsymbol{k}_\perp)\ ,
\label{eq:bperp}
\end{equation}
from which an ensemble average for $\lan b_\perp \ran = \lan |{\boldsymbol b}_\perp| \ran$ may be computed analogously to Eq.~(\ref{eq:kt1}).
%

%
%
%
\subsection{The $\sigma_{c\overline{c}}$ calculation}
\label{ssec:sigmodel}
Physically, the expression in Eq.~(\ref{eq:sigdef}) comes from the formal definition of the nucleon mass connecting it to the
trace of the QCD energy-momentum tensor \cite{Shifman:1978zn},
\begin{align}
M\, \overline{\psi}_N \psi_N\ =\ \lan p\, \big|\, &\theta_{\mu\mu}\, \big| p \ran \\
&\theta_{\mu\mu}\ \equiv\ \dots\, + \sum_{Q=(c,b,t)}\! m_Q\, \overline{Q} Q\, + \dots\ , \nonumber
\end{align}
in which the `dots' above contain contributions from light quarks and glue as well as higher tensor components extraneous
to this analysis. The contribution from a specific heavy quark flavor to the total mass of the nucleon can then be computed as
a scalar form factor \cite{Gasser:1990ce,Gasser:1990ap} at zero momentum transfer as depicted in Fig.~\ref{fig:diagram}({\bf b}),
in which the external scalar field couples to the internal quark lines with an interaction strength
equal to the heavy quark mass.

We therefore compute the relevant matrix element as
\begin{align}
2M\,\sigc\ &=\ \bar{u}(p)\, \Big[ \sigma^\circ_c\, \big(q^2\!=\!0\big)\, +\, \sigma^\circ_{\bar{c}}\, \big(q^2\!=\!0\big)  \Big]\, u(p)\ ;
\label{eq:sigFORM}
\end{align}
taking $\sigma^\circ_c$ above from the amputated diagram of Fig.~\ref{fig:diagram}({\bf b}) such that
$\sigma_c \equiv \overline{u}\, \sigma^\circ_c\, u \big/ (2M)$, we have
\begin{align}
\sigma_c\ &=\ {ig_c \over 2M}\, \int {d^4k \over (2\pi)^4}\ \overline{u}(p) \left( 1 \over \ksl - m_c + i \epsilon \right) \big[ m_c \mathcal{I}_4 \big]
\left( 1 \over \ksl - m_c + i \epsilon \right)\, u(p)\ \left( { 1 \over [p-k]^2 - M^2_S + i\epsilon } \right) \nonumber \\
&\hspace{2.95cm} \times\, \left( { \Lambda^2 \over k^2 - \Lambda^2_c + i\epsilon } \right)^{2 \gamma}\ ,
\label{eq:sigCOV}
\end{align}
where we have incorporated the covariant quark-nucleon vertex function $\phi_c$ of Eq.~(\ref{eq:phi}) on the second line, noting $t_c = k^2$. 
Assuming $m_{\bar{c}} = m_c$, for the covariant calculation with $\gamma = 3$ in the $t_c$-dependent quark-nucleon
vertex function we may evaluate the expression in Eq.~(\ref{eq:sigCOV}) using standard techniques --- shifting
the loop momentum $k \to \ell = k - zp$ and introducing Feynman parameters:
\begin{equation}
\sigma_c\ =\ { i g_c m_c \Lambda^{4\gamma}_c }\, \left({ \Gamma{[3+2\gamma]} \over \Gamma{[2\gamma]} } \right)
\int\! dx\, dy\, dz\ \delta{\big( 1 - [x+y+z] \big)}\, x z^{2\gamma - 1}
\int\! {d^4\ell \over (2\pi)^4}
\left\{ { \ell^2 + \big( m_c + yM \big){}^2 \over \big( \ell^2 - \delta + i\epsilon \big){}^{3+2\gamma} } \right\}\ .
\label{eq:sig_gengam}
\end{equation}
Then, specializing to the main prescription used in this analysis, $\gamma = 3$ for a $4$-quark spectator, we
arrive at the covariant expressions to be evaluated in terms of the model parameters defined in Sect.~\ref{ssec:DIS},
{\it viz.}
\begin{align}
\label{eq:sigc}
\sigma_c\ &=\ 2 g_c m_c\, \left({\Lambda^6_c \over 4\pi}\right)^2 \int_0^1 dx\, \int_0^{1-x} dy\ x (1-x-y)^5
\left\{ -\!\left({1 \over \delta} \right)^6\, +\, 3\, [ m_c + y M ]^2 \left({1 \over \delta} \right)^7 \right\}  \\
\sigma_{\bar{c}}\ &=\ 2 g_{\bar{c}} m_c\, \left({\Lambda^6_{\bar{c}} \over 4\pi}\right)^2 \int_0^1 dx\, \int_0^{1-x} dy\ x (1-x-y)^5
\left\{ -\!\left({1 \over \overline{\delta}} \right)^6\, +\, 3\, [ m_c + y M ]^2 \left({1 \over \overline{\delta}} \right)^7 \right\}\ ,
\label{eq:sigcb}
\end{align}
where the squared mass parameters $\delta$ and $\overline{\delta}$ appearing in Eqs.~(\ref{eq:sig_gengam})--(\ref{eq:sigcb}) are explicitly given by
\begin{align}
\delta\ &\equiv\ x\, m^2_c + y\, M^2_S + (1-x-y)\, \Lambda^2_c - y(1-y)\, M^2 \nonumber \\
\overline{\delta}\ &\equiv\ x\, m^2_c + y\, M^2_{\bar{S}} + (1-x-y)\, \Lambda^2_{\bar{c}} - y(1-y)\, M^2\ ,
\end{align}
and the full result for $\sigma_{c\overline{c}}$ is a sum of these,
\begin{equation}
\sigma_{c\overline{c}}\ =\ \sigma_c\, +\, \sigma_{\bar{c}}\ .
\label{eq:sigfin}
\end{equation}
We explore the range of possibilities for this latter quantity in the next section after constraining from
hypothetical DIS information the adjustable parameters in Eqs.~(\ref{eq:sigc})--(\ref{eq:sigfin}).
%
%
%
%
\section{MCMC simulations}
\label{sec:MCMC}
\paragraph{Statistical methodology.}
To test quantitatively the possible connection between $\fcc$ and $\sigc$, a systematic exploration of
the parameter space established in Sect.~\ref{sec:LF} is unavoidable. Traditional model-fitting methods using gradient minimization
or Hessian techniques tend to be overly sensitive to initial conditions and generally require {\it a priori} knowledge
of the parameter space to avoid local critical points while searching for global minima in $\chi^2$. An alternative approach
that affords a wider and more systematic exploration of the parametric dependence of the model for $\fcc$ and $\sigc$ is MCMC
\cite{Robert:2005} --- a statistical sampling method for Bayesian parameter inference. (Brief statistical reviews may be found
in Chapts.~$39$, $40$, and the associated references of Ref.~\cite{Olive:2016xmw}.)
MCMC statistical analyses operate by performing large samplings of the multi-dimensional parameter space
by effectively simulating a model $n_{sim}$ times and thereby probing its {\it joint posterior distribution},
\begin{align}
p(\vec{\theta}\, \big| \xi )\ \sim\ &p(\xi \big| \vec{\theta}\, )\, p(\vec{\theta}\,) \nonumber\\
&p(\xi \big| \vec{\theta}\, )\ =\ \exp{\big(\! -\!\chi^2 \big/ 2 \big)}\ ,
\label{eq:Bayes}
\end{align}
an object that quantifies the probability of a specific combination of model parameters ($\vec{\theta}\,$)
given some collection of input data ($\xi$). The {\it likelihood function} $p(\xi \big| \vec{\theta}\,)$ defined above is itself a function
of the conventional $\chi^2$, and $p(\vec{\theta}\,)$ contains the {\it prior} distributions of model parameters.
The joint probability distribution sampled in MCMC calculations is a large multi-dimensional object, and
MCMC runs thus entail the production of an expansive chain (having length $n_{sim}$) of parameter values; these chains represent
the favored loci of greatest probability density in the space spanned by $\vec{\theta}$, subject to the constraints imposed by data
through the likelihood function of Eq.~(\ref{eq:Bayes}).
Runs may ultimately yield histograms proportional to the probability distribution functions (p.d.f.s) for quantities
of interest like the $\sigc$ matrix element after binning the large ensemble of values that result from evaluating these
quantities according to the MCMC chain.
In practice, we make use of the delayed-rejection, adaptive Metropolis (DRAM) {\small FORTRAN} algorithms of Haario {\it et al.} \cite{Laine} to explore
the parameter space of the model described in Sect.~\ref{sec:LF}. Taking broad Gaussian priors for model parameters, MCMC algorithms
scan the $6$-dimensional parameter space, informing the likelihood function of Eq.~(\ref{eq:Bayes}) with pseudo-data in the form of $4$ evenly spaced
$x$-points of $x(c + \overline{c})$ over $0.2 \le x \le 0.5$ as given by the ``confining'' model of
Ref.~\cite{Hobbs:2013bia}, plotted in Fig.~\ref{fig:LF-model}. We set overall normalizations to give
$\lan x \ran_{c+\overline{c}} = 0.001 \pm 50\%,\, 0.0035 \pm 50\%$, and $0.01 \pm 50\%$ --- corresponding respectively in the first
two cases to the upper limit determined in the full QCD global analysis of Ref.~\cite{Jimenez-Delgado:2014zga}
for the standard high-energy data set ({\it i.e.}, excluding the EMC measurements of Ref.~\cite{Aubert:1982tt}) 
as well as the value preferred by EMC measurements when fitted in isolation, as also done in
Ref.~\cite{Jimenez-Delgado:2014zga}. We note as well that this work found a $4\sigma$ limit of
$\lan x \ran_{c+\overline{c}} \le 0.005$ under an alternative choice for the charm threshold suppression factor used in the
global analysis. The intermediate IC magnitude used here, $\lan x \ran_{c+\overline{c}} = 0.0035$, is also comparable to the central value(s)
found in the most recent NNPDF3.1 analysis \cite{Rojo:2017xpe} of intrinsic charm. Lastly, the biggest hypothetical IC magnitude,
$\lan x \ran_{c+\overline{c}} = 0.01 \pm 50\%$, corresponds to larger limits found in some other recent works ({\it e.g.}, the 2017
CTEQ-TEA IC analysis \cite{Hou:2017khm}) that imposed more strongly restrictive cuts in $Q^2$ and $W^2$ than those used in
Refs.~\cite{Jimenez-Delgado:2014zga,Jimenez-Delgado:2015tma}.
Performing the aforementioned $6$-dimensional scans of the parameter space, we allow
the mass of the charm quark to vary about a Gaussian input prior centered at $m_c = m_{\bar{c}} = 1.3$ GeV, and
sample the model space $n_{sim} = 10^5$ times according to a likelihood function constrained to the input pseudo-data
shown in Fig.~\ref{fig:LF-model}. This procedure results in the desired MCMC chain from which we may use a randomly drawn subset of parameter values
to compute $\fcc$, which yields the gray curves plotted next to the input pseudo-data in Fig.~\ref{fig:LF-model}.
Given that the charm PDF $c(x)$ in Eq.~(\ref{eq:cx}) depends on the charm quark-spectator invariant mass $s_{cS}(x,k^2_\perp)$
and is dominated by small $k^2_\perp \sim 0$ by merit of the vertex function $\phi_c(x,k^2_\perp)$, the distributions predicted
by Eq.~(\ref{eq:cx}) generally peak at
\begin{equation}
x_{\mathrm{peak}}\ \approx\ { m_c \over  m_c + M_{S/\bar{S}} }\ \simeq\ 0.2 - 0.3\ ,
\end{equation}
in rough agreement with the gray curves of Fig.~\ref{fig:LF-model}.
As a further component of the statistical analysis, we exhibit in Fig.~\ref{fig:2D} the joint posterior distribution of
Eq.~(\ref{eq:Bayes}) as probed by MCMC techniques. This is accomplished by binning the MCMC chain in each of the parameters
(thereby effectively marginalizing the others), resulting in the diagonal entries of Fig.~\ref{fig:2D}, which display histograms
proportional to the posterior probability distribution functions (p.d.f.s)~for $g_c, m_c, \Lambda_c, \Lambda_{\bar{c}}, M_S$,
and $M_{\bar{S}}$, respectively. At the same time, we also plot the two-dimensional correlations of the various parameters
amongst one another, yielding the off-diagonal panels of Fig.~\ref{fig:2D}. In each case this is done for two assumed strengths
in Eq.~(\ref{eq:phi}) --- the main prescription of this work, $\gamma = 3$ (black histogram curves and correlations {\it above}
the diagonal), and the weaker form with $\gamma = 1$ (blue histogram curves and correlations {\it below} the diagonal). The
qualitative similarity of the parametric behavior seen for these different choices comports with the independence
of the main results of this paper from the value of $\gamma$ used in Eq.~(\ref{eq:phi}).
%

%
\begin{figure}
\hspace{-0.5cm}
\includegraphics[scale=0.32]{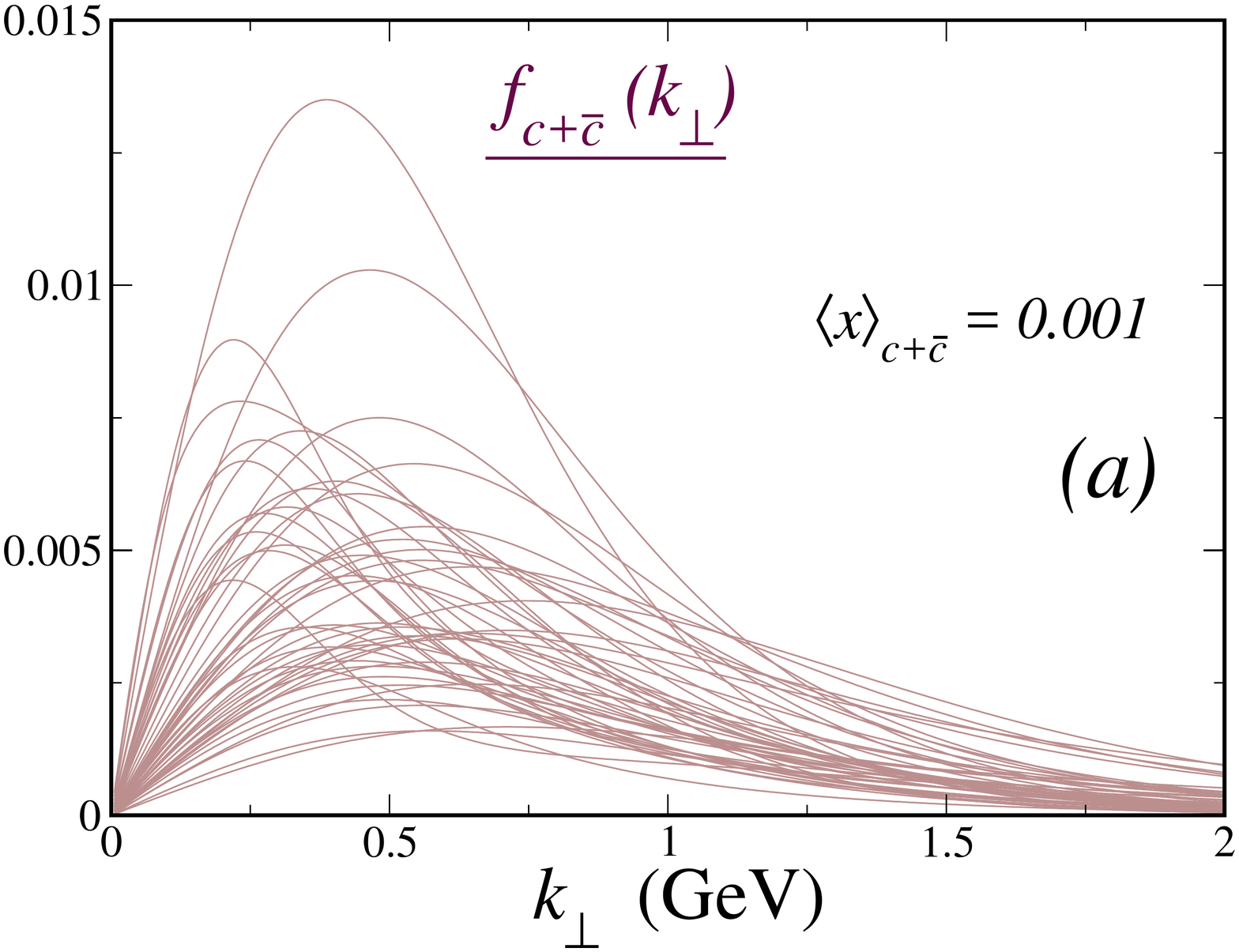} \ \ \
\includegraphics[scale=0.32]{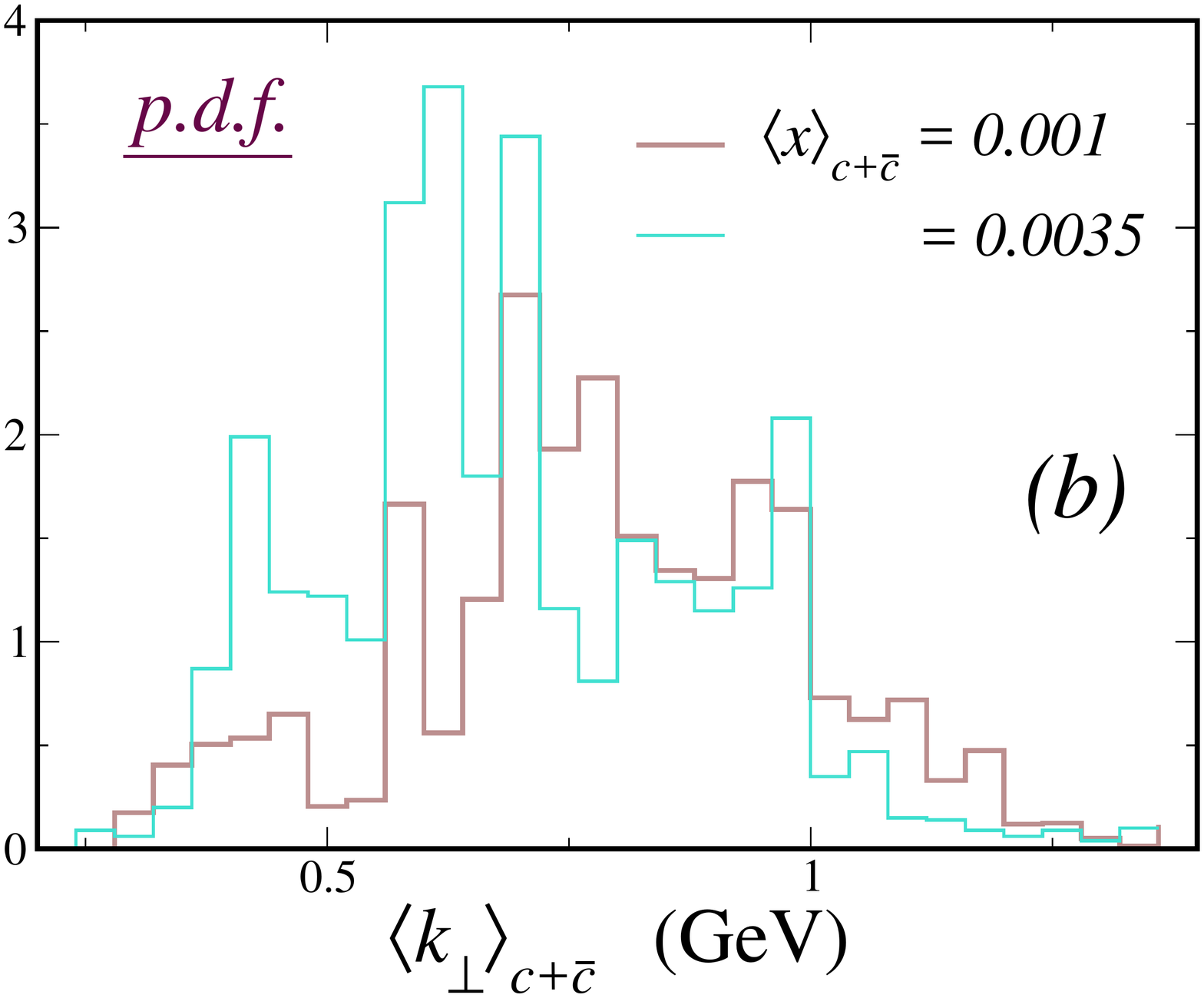}
\caption{
(Color online) Analogous to the plots of Fig.~\ref{fig:LF-model}, selecting a random ensemble of parameter
combinations from the full MCMC chain allows a sample collection of DIS-constrained $k_\perp$ distributions
\`a la Eq.~(\ref{eq:TMD}) to be plotted ({\bf a}). We show a random subset of curves generated using the
$t_c$-dependent $\phi_c$ vertex function with $\gamma=3$ in the $\lan x \ran_{c+\overline{c}} = 0.001$ scenario; the
complementary sets for $\lan x \ran_{c+\overline{c}} = 0.0035$ and $0.01$ are virtually indistinguishable as attested by the associated
p.d.f.s for the average transverse momentum $\lan k_\perp \ran_{c+\overline{c}}$ of Eqs.~(\ref{eq:kt1})--(\ref{eq:kt2}) in the
scalar spectator model ({\bf b}). This latter quantity has the interpretation of being the mean $\lan k_\perp \ran_{c+\overline{c}}$
of the distributions plotted in ({\bf a}). Here we plot p.d.f.s for the $\gamma=3$ interaction of Eq.~(\ref{eq:phi})
for both smaller IC normalizations, $\lan x \ran_{c+\overline{c}} = 0.001$ (heavy linestyles) and $\lan x \ran_{c+\overline{c}} = 0.0035$
(thin linestyles).
}
\label{fig:kT}
\end{figure}

\paragraph{$k_\perp$ dependence.}
As one additional check of the model we benchmark MCMC output by computing histograms of the average quark
transverse momentum $\lan k_\perp \ran_{c+\overline{c}}$ defined in Eqs.~(\ref{eq:kt1})--(\ref{eq:kt2}) of
Sect.~\ref{ssec:DIS}.
The average $\lan k_\perp \ran_{c+\overline{c}}$ is directly computable from TMD distributions obtained according
to Eq.~(\ref{eq:TMD}), and for the normalization $\lan x \ran_{c+\overline{c}} = 0.001$ we show a representative
ensemble of $f_{c+\overline{c}}(k_\perp) = [f_{c}(k_\perp) + f_{\bar{c}}(k_\perp)] \big/ 2$ in Fig.~\ref{fig:kT}({\bf a}).
From the TMDs, we determine histograms for $\lan k_\perp \ran_{c+\overline{c}}$ itself and display the results for the two lesser
normalizations $\lan x \ran_{c+\overline{c}} = 0.001$ and $0.0035$ in Fig.~\ref{fig:kT}({\bf b}), where the curves again correspond
to the $\gamma=3$ interaction of Eq.~(\ref{eq:phi}). Taking the mean and statistical error (the latter from the standard deviation)
for these we get
\begin{align}
\lan k_\perp \ran_{c+\overline{c}}\ &=\ 0.73 \pm 0.20\, \mathrm{GeV} \hspace*{2cm} \lan x \ran_{c+\overline{c}} = 0.001 \nonumber \\
                   &=\ 0.76 \pm 0.22\, \mathrm{GeV} \hspace*{2cm} \lan x \ran_{c+\overline{c}} = 0.0035\ .
\end{align}
As such, the covariant power-law interaction constrained to DIS pseudo-data prefers an average transverse
momentum for the charm pair much larger than a typical light-quark scale \cite{Hobbs:2016xfz},
$\lan k_\perp \ran_{c+\overline{c}} \gg 200\, \mathrm{MeV}$, corresponding to deeply internal configurations with
small transverse impact parameters $\lan b_\perp \ran_{c+\overline{c}} \ll 1\, \mathit{fm}$ \`a la Eq.~(\ref{eq:bperp}).
We observe that the mean $\lan k_\perp \ran_{c+\overline{c}}$ is largely independent of the magnitude of
$\lan x \ran_{c+\overline{c}}$ (similar results obtain for the largest normalization $= 0.01$) and is
more directly controlled by the IC distribution shape.
%

%
\begin{figure}
\includegraphics[scale=0.31]{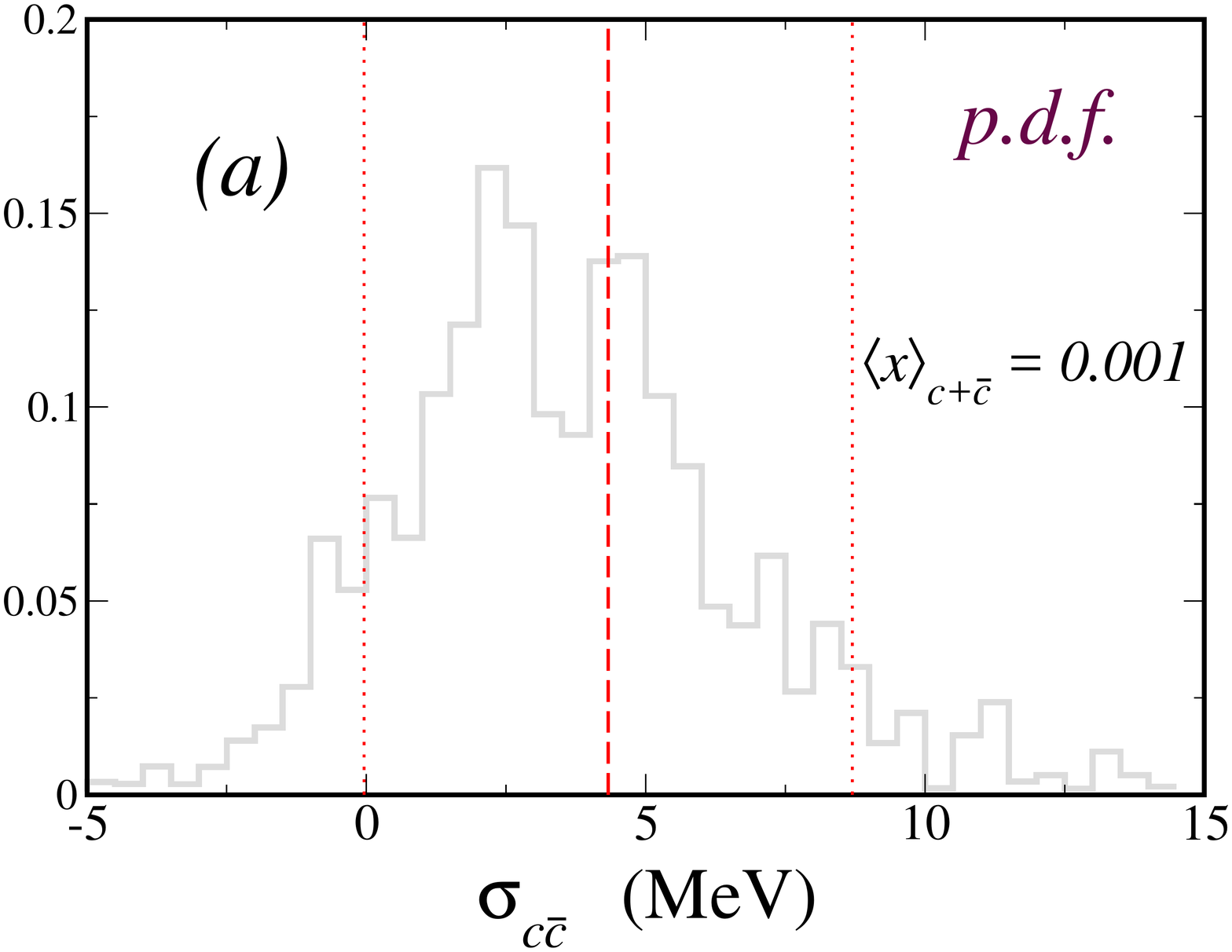} \ \ \ \ \
\raisebox{0cm}{\includegraphics[scale=0.31]{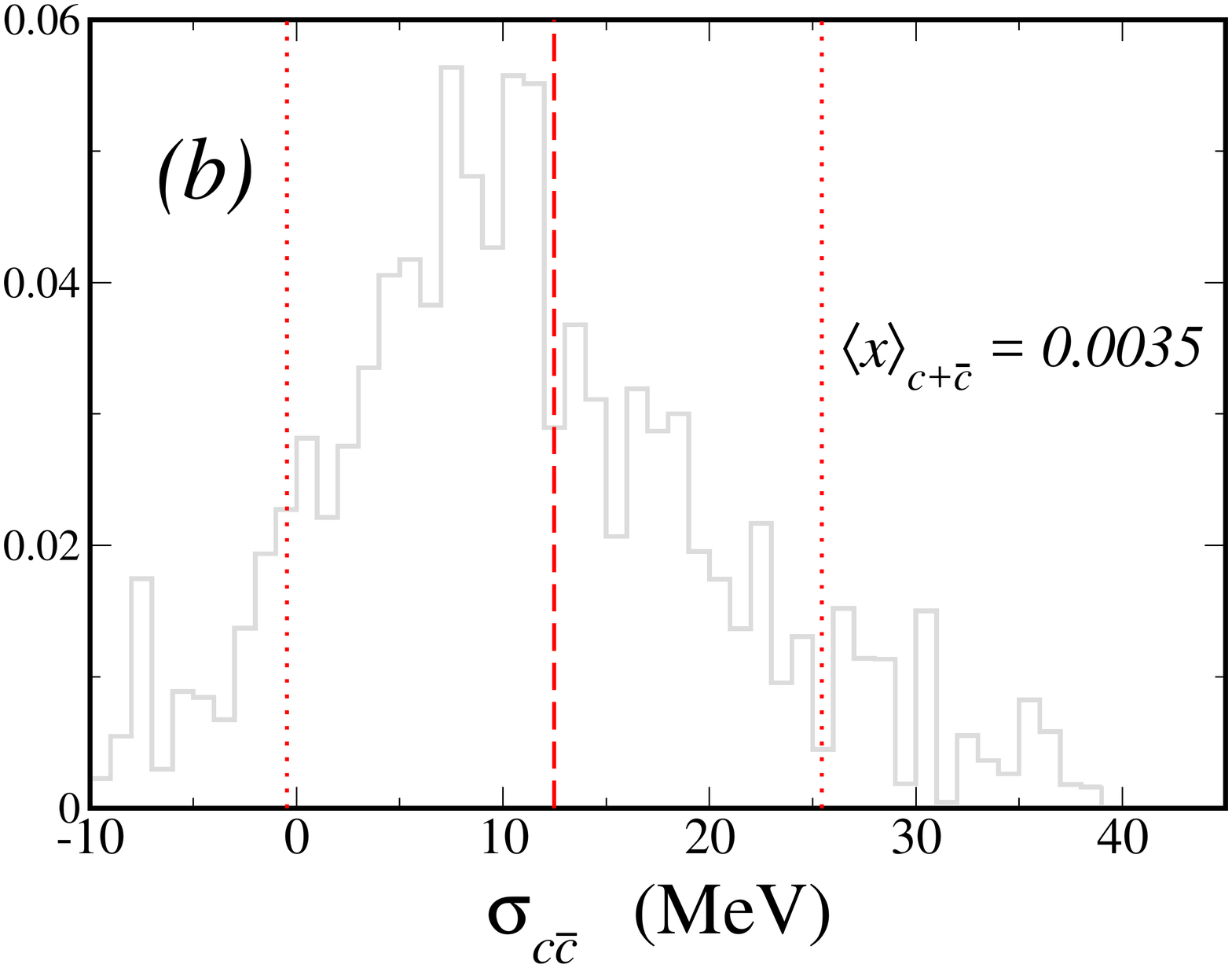}} \ \ \ \ \
\raisebox{0cm}{\includegraphics[scale=0.31]{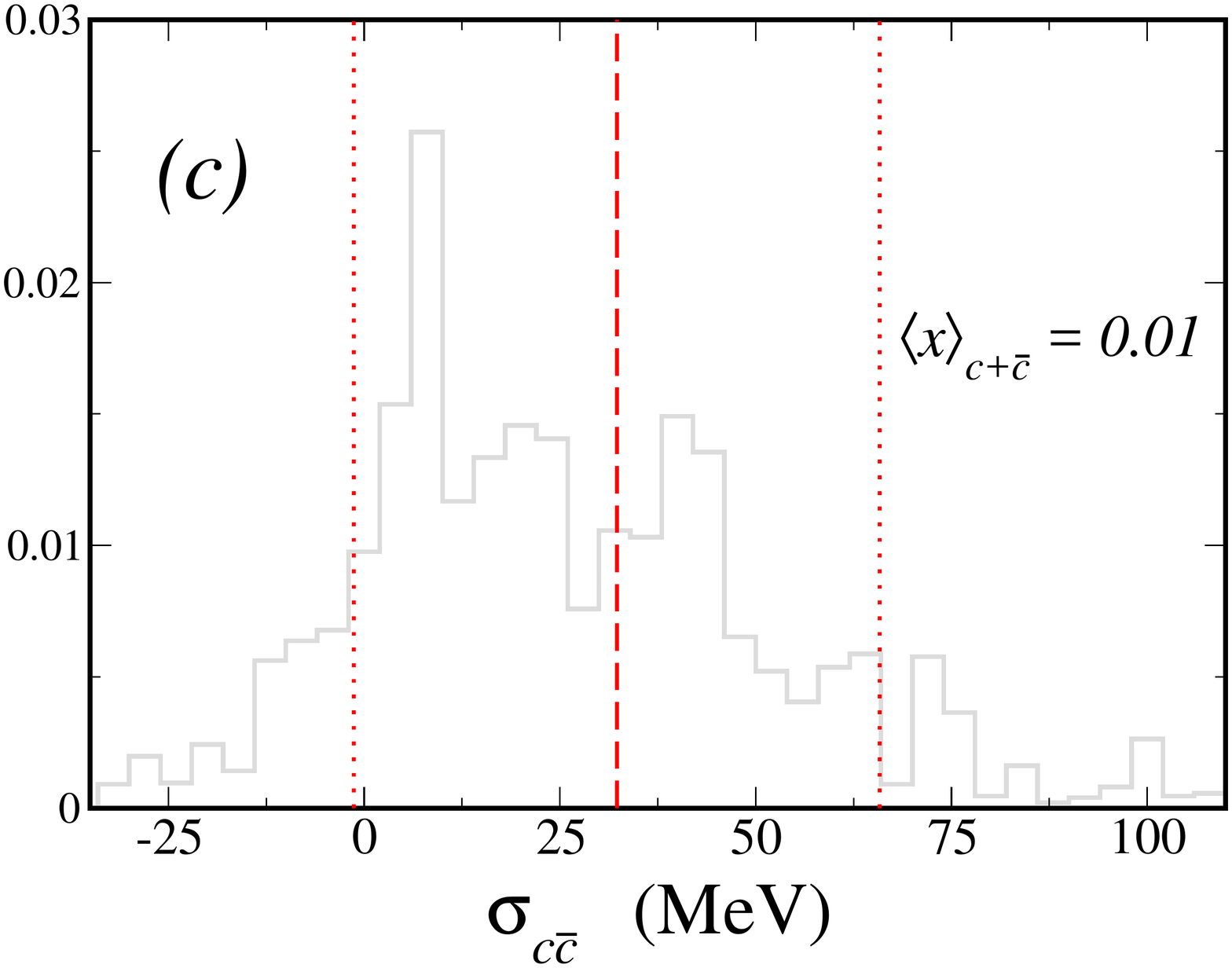}}
\caption{(Color online) ({\bf a}) The p.d.f.~for $\sigc$ as determined from an MCMC chain
constrained by pseudo-data with a normalization consistent with the most conservative upper limit in Ref.~\cite{Jimenez-Delgado:2014zga},
$\lan x \ran_{c+\overline{c}} = 0.001$; scans were performed using $\gamma=3$ in the covariant quark-nucleon interaction of Eq.~(\ref{eq:phi}). We
also show the results of the same calculation using input pseudo-data with a larger overall normalization consistent with the separately fitted EMC data
--- namely, $\lan x \ran_{c+\overline{c}} = 0.0035$ ({\bf b}) --- as well as a still greater normalization $\lan x \ran_{c+\overline{c}} = 0.01$
({\bf c}). There is a close similarity between the p.d.f.s obtained for different assumed strengths of the power-law vertex function of Eq.~(\ref{eq:phi}),
and we only plot results following from the $\gamma=3$ scenario on the basis of counting rules here. Specific numerical values are described in the
accompanying text.
}
\label{fig:hist}
\end{figure}
\begin{figure}
{\includegraphics[scale=0.45]{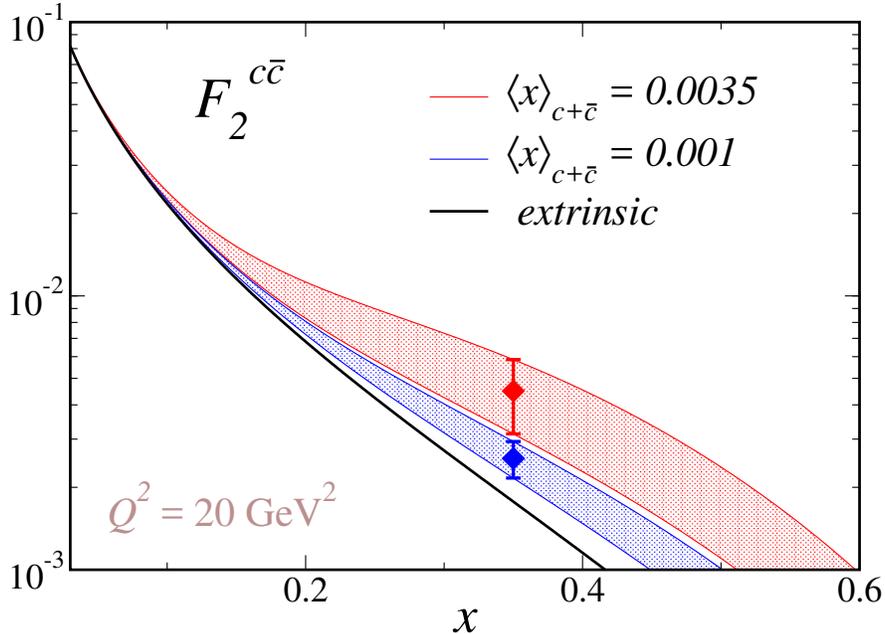}}
\caption{
(Color online) The charm structure function $\fcc(x,Q^2)$ evolved by massless DGLAP to a scale accessible to the MEIC proposal of
Ref.~\cite{Accardi:2012qut}, $Q^2 = 20$ GeV$^2$. The solid black line is the
extrinsic contribution from NLO evolution, while the lower and upper bands correspond to the two IC scenarios of lesser magnitude
investigated here, $\lan x \ran_{c+\overline{c}} = 0.001$ and $0.0035$ (at $Q^2=m^2_c$), respectively. While the vertical scale suggests the high experimental
precision that would be necessary to unambiguously resolve IC from zero at these smaller magnitudes, the structure function corresponding
to the upper-level IC scenario $\lan x \ran_{c+\overline{c}} = 0.01$ explored in Fig.~\ref{fig:hist} would overhang these.
}
\label{fig:Q20}
\end{figure}

\paragraph{$\sigc$ results.}
We now turn the MCMC chains described above to the computation of $\sigc$.
In general, the $\sigc$ histograms obtained from DRAM scans of the $6$-parameter model space are well-described with simple
normal distributions, although for simplicity and directness we ultimately interpolate and plot the resulting p.d.f.s~in Fig.~\ref{fig:hist},
which shows the main results of the present analysis. Corresponding to input pseudo-data such as is shown in Fig.~\ref{fig:LF-model} for the two
smaller IC normalizations, the panels of Fig.~\ref{fig:hist} display the
$\sigc$ p.d.f.~for the covariant $\gamma=3$ quark-nucleon interaction of Eq.~(\ref{eq:phi}) under each of our three hypothetical scenarios
for the IC magnitude --- $\lan x \ran_{c+\overline{c}} = 0.001$ ({\bf a}),
$\lan x \ran_{c+\overline{c}} = 0.0035$ ({\bf b}), and $\lan x \ran_{c+\overline{c}} = 0.01$ ({\bf c}). With these results,
we may extract the constraints placed on the size of the nucleon's charm sigma term by hypothetical information
on $\fcc$ in the context of the covariant quark $+$ spectator model.
Based upon the distribution plotted in panel ({\bf a}) of Fig.~\ref{fig:hist}, we conclude the $1\mathrm{C.L.} = 68\%$ limit on $\sigma_{c\overline{c}}$
corresponding to $\lan x \ran_{c+\overline{c}} = 0.001$ with the quark-nucleon interaction choice of Eq.~(\ref{eq:phi}) to be
\begin{align}
\sigma_{c\overline{c}}\ =\ 4.3 \pm 4.4\, \mathrm{MeV} \hspace*{2cm} \big(\mathrm{\gamma=3,}\ \ \lan x \ran_{c+\overline{c}} = 0.001\big)
\end{align}
in keeping with the shallower overall normalization of the charm quark wave function in this case.
On the other hand, if MCMC runs are informed by pseudo-data with bigger normalizations
consistent with $\lan x \ran_{c+\overline{c}} = 0.0035$ as inspired by the EMC data fitted apart from the standard QCD set
in \cite{Jimenez-Delgado:2014zga}, we obtain larger $1\sigma$ limits roughly compatible with the lower bounds of some lattice
studies,
\begin{align}
\sigma_{c\overline{c}}\ =\ 12.5 \pm 13.0\, \mathrm{MeV} \hspace*{2cm} \big(\mathrm{\gamma=3,}\ \ \lan x \ran_{c+\overline{c}} = 0.0035\big)\ ,
\end{align}
which is associated with the p.d.f.~shown in Fig.~\ref{fig:hist}({\bf b}).
Using our largest overall input normalization $\lan x \ran_{c+\overline{c}} = 0.01$ as is allowed, for instance, by the most recent CTEQ-TEA
analysis, we find a still wider range at the $1\sigma$ level,
\begin{align}
\sigma_{c\overline{c}}\ =\ 32.3 \pm 33.6\, \mathrm{MeV} \hspace*{2cm} \big(\mathrm{\gamma=3,}\ \ \lan x \ran_{c+\overline{c}} = 0.01\big)\ ,
\label{eq:sig01}
\end{align}
corresponding to a band having significant overlap with the lattice \cite{Freeman:2012ry,Gong:2013vja,Abdel-Rehim:2016won} and N$^3$LO pQCD 
\cite{Kryjevski:2003mh} determinations. This maximal scenario for $\sigc$ is illustrated in Fig.~\ref{fig:hist}({\bf c}).
Again we note that the choice of parametric form for the relativistic vertex function has little effect on the center and shape of the resulting $\sigc$
probability distributions plotted in Fig.~\ref{fig:hist}, and the dominant factor influencing the preferred value for the sigma
term is the overall magnitude of the nonperturbative charm component, $\lan x \ran_{c+\overline{c}}$. This is
confirmed by separate calculations using different sizes for the pseudo-data error bars and different assumed shapes for the input
distributions $x(c+\overline{c})$, all of which return qualitatively similar results. We conclude the connection between the total
magnitude of the valence-like intrinsic charm and model prediction for $\sigc$ to be a solid feature of the nonperturbative wave
function and independent of the prescription for the quark-nucleon vertex used to regulate the ultraviolet behavior of the model.
The existence of this connection is the main conclusion of the present article, and in the absence of unambiguous data requires the
assumption of several pseudo-data points with a generic shape predicted by various models \cite{Hobbs:2013bia} favoring valence-like IC distributions. For
definiteness it was necessary to take specific uncertainties and $x$-values for these input points, and the resulting detailed dependence
of the predicted $\sigc$ on this input pseudo-data is non-trivial, the observations of the last paragraph notwithstanding. For instance,
while reducing the overall size of the relative uncertainties ({\it e.g.}, from $50\%$ to $25\%$) on the pseudo-data points shown in
Fig.~\ref{fig:LF-model} generally produces some
marginal tightening in the predicted ranges for $\sigc$ binned in Fig.~\ref{fig:hist}, we find that additional input points at higher $x$ can
potentially have a more sizable effect. As an illustration, we find constraining the parameter space with an additional point at $x=0.6$
(also taken from the confining model of Ref.~\cite{Hobbs:2013bia} with $\lan x \ran_{c+\overline{c}} = 0.01$) modifies the predicted
charm sigma term in Eq.~(\ref{eq:sig01}) as $\sigma_{c\overline{c}} = 32 \pm 34\, \mathrm{MeV} \to\ 44 \pm 38\, \mathrm{MeV}$.
These observations add to the already urgent need to more carefully measure the charm structure function at high $x$ and
intermediate $Q^2$ as might be carried out at a future electron-ion collider (EIC) \cite{Accardi:2012qut}. From
exploratory estimates of the kinematical coverage of the $\sqrt{s} = 45$ GeV MEIC-like proposal, for instance,
at a favorable scale of \linebreak $\lan Q^2 \ran \sim 20$ GeV$^2$ access to bins at reasonably high $x$, $0.3 \le x \le 0.4$, should be possible,
albeit with somewhat reduced luminosity. We plot the two lower-magnitude IC scenarios evolved \cite{Miyama:1995bd} using massless DGLAP at NLO
to this proposed scale in Fig.~\ref{fig:Q20}, showing the precise measurement that would be required to distinguish the nonperturbative
part of the charm structure function at this level. This calculation illustrates the approximate magnitude expected for the charm
structure function's IC component, and describing its full scale dependence requires a more involved pQCD treatment \cite{Hoffmann:1983ah}.
Additional complementary measurements might also be performed at LHCb \cite{Boettcher:2015sqn} or AFTER@CERN \cite{Brodsky:2012vg},
where precise extractions of charmed hadron rapidity distributions might further constrain the charm PDFs and provide indirect information that
could similarly tighten knowledge of $\sigc$ along the lines described here.
%
%
%
%
\section{Conclusions}
\label{sec:conc}
The foregoing analysis represents a first effort to simultaneously model the influence of nonperturbative charm
in both the nucleon's structure function and sigma term. We have formulated a simple constituent quark model with enough flexibility
to describe a wide range of IC PDF shapes and magnitudes, by which we conclude that there is a direct proportionality between the size of
IC as extracted in DIS in terms of the charm momentum fraction $\lan x \ran_{c+\overline{c}}$ and the magnitude of the charm sigma term
$\sigc$. We have established this connection by performing unbiased MCMC scans of the $6$-dimensional parameter space of the quark $+$
scalar model developed in Sect.~\ref{sec:LF}, subject only to input pseudo-data proportional to the nonperturbative charm structure
function at partonic threshold, {\it i.e.}, $\fcc(x,Q^2=m^2_c)$; these inputs were taken to conform to the valence-like
$x$ dependence established in earlier models \cite{Brodsky:1980pb,Hobbs:2013bia} with an overall normalization tied to the predictions
or limits of recent QCD global analyses. On this basis, we find that steadily larger sizes for $\sigc$ are tolerated with successively
larger input normalizations for $\lan x \ran_{c+\overline{c}}$. As such, the smallest IC scenario considered, 
$\lan x \ran_{c+\overline{c}} = 0.001$ corresponds to a minimal predicted charm sigma term $\sigc \sim 4$ MeV peripheral to the
typical value(s) preferred by recent lattice calculations. On the other hand, calculations with larger IC normalizations, 
$\lan x \ran_{c+\overline{c}} = 0.0035\, , 0.01$ approach and/or agree with the few current lattice results at the $1\sigma$
level.
%

%
Despite this rough potential consistency with extant lattice calculations, we note that model calculations here generally tend to
prefer somewhat lower magnitudes for $\sigc$, with the implication for the extended electroweak doublet BSM scenario of
Ref.~\cite{Hill:2013hoa} being a slight favorability for the definite prediction of the WIMP-nucleon cross section
$\sigma_{\mathrm{SI}} \sim 10^{-49}$ at smaller $\sigc$.
Notably, we find that several points of input pseudo-data at intermediate $x$ for the combination $\fcc$ tend to lead to
MCMC simulations that do not necessarily strongly distinguish $\sigc$ from zero. Rather, to resolve non-zero $\sigc$ would seem to
require still more information in the form of additional measurements, especially at higher $x$. Were it possible to extract
unambiguously, other specifically nonperturbative information, such as precise values for the charge asymmetry
\begin{equation}
C^- = \int_0^1 dx\ x\, \Big( c(x) - \overline{c}(x) \Big)
\label{eq:casym}
\end{equation}
might complement better knowledge of the $C$-even $F^{c\overline{c}}_2$ structure function and have the potential to more strongly limit
the parameter space and narrow the allowed range for $\sigc$.
A fuller calculation would allow vector configurations for the $4$-quark spectator state and relax the static mass assumption for
the constituent particles used here. More detailed modeling along these lines, however, would likely require additional empirical information
to meaningfully constrain the more expansive parameter space. We also disclaim the fact that the constituent quark model itself, while
well-adapted to describe DIS structure functions, may turn out to be more approximative in treating $\sigc$, which as the matrix element
of a different operator may differently weigh the dynamical effects at work in the quark-nucleon interaction. This possibility might
be assessed and controlled in part by extending the present calculation to the PDFs, form factors, and sigma terms of other
flavors with different spin-flavor configurations to further validate the approach here. While beyond the scope of the present article,
such work could provide both the setting and means for making such refinements.
In any event, the unambiguous message of this analysis is the importance of new empirical information directly bearing on the charm PDF,
with the understanding for experimental efforts that observations of $\fcc$ at large $x$ and intermediate $Q^2$ is most likely
to have the greatest impact as noted in Sect.~\ref{sec:MCMC}. A future EIC \cite{Accardi:2012qut} would be uniquely well-disposed to
carry out measurements of this sort. Hadronic collisions might also yield complementary information on the charm
PDF at high $x$ through channels like weak boson production in coincidence with a charm jet \cite{Boettcher:2015sqn,Brodsky:2012vg}, {\it i.e.},
$pp \to c + Z/W^{\pm} + X$.
Moreover, with very high precision it may eventually become possible to observe or constrain the charm asymmetry $C^-$ of
Eq.~(\ref{eq:casym}) from differences in rapidity distributions for the hadroproduction of electroweak bosons
\cite{Brodsky:2015fna}.
Alongside these necessary experimental developments, more and improved lattice calculations of $\sigc$ can bring clarity to the IC
question. On the basis of this work we find that a potential reciprocity may exist between efforts to isolate an IC PDF from the world's
data on QCD processes and lattice and other calculations of $\sigc$. It may be possible in the future, for instance, to construct
still more comprehensive global analyses on the basis of this reciprocity to impose constraints on PDF fits from lattice determinations
of the sigma term. Doing so has the potential to open new fronts and aid in the resolution of the nucleon's heavy quark structure.
%
%
%
\section*{Acknowledgements}
We are grateful to Andr\'e Walker-Loud and Richard Hill for initial discussions motivating this work. For helpful conversations
regarding IC in DIS we thank Wally Melnitchouk, Pavel Nadolsky, Jean-Philippe Lansberg, and Stan Brodsky. We also acknowledge
Marko Laine for helpful correspondence on the DRAM numerical routines as well as Scott Pratt, Paul Desmedt, and Kris Thielemans
for exchanges regarding Bayesian analyses; we thank Xilin Zhang for input regarding statistical
calculations and for suggesting our consideration of $\lan k_\perp \ran_{c+\overline{c}}$.
%
%
%
\section*{References}
\end{document}